\documentclass[aps,amssymb,amsmath,reprint]{revtex4-1} 

\usepackage{graphicx}  
\usepackage{dcolumn}   
\usepackage{physics}
\usepackage{color}
\usepackage{appendix}
\usepackage[normalem]{ulem}
\usepackage{lineno}
\definecolor{orange}{rgb}{0.8, 0.3, 0}

\definecolor{blueviolet}{rgb}{0.2, 0.2, 0.6}
\usepackage[pdftex, 
bookmarks=true, 
colorlinks=true,
allcolors=blueviolet,
pdfstartview={FitH}, 
]{hyperref}

\begin{document}

\title{In-operando microwave scattering-parameter calibrated measurement of a Josephson travelling wave parametric amplifier}

\author{S.-H. Shin$^{1,2}$}
\author{M. Stanley$^{1}$}
\author{W. N. Wong$^{1}$}
\author{T. Sweetnam$^{1}$}
\author{A. Elarabi$^{1}$}
\author{T. Lindstr\"om$^{1}$}
\author{N. M. Ridler$^{1}$}
\author{S. E. de Graaf$^{1}$}
\email{sdg@npl.co.uk}

\affiliation{$^1$National Physical Laboratory, Teddington TW11 0LW, United Kingdom}
\affiliation{$^2$Sejong University, Seoul 05006, Republic of Korea}

\begin{abstract}
Superconducting travelling wave parametric amplifiers (TWPAs) are broadband near-quantum limited microwave amplifiers commonly used for qubit readout and a wide range of other applications in quantum technologies. The performance of these amplifiers  depends on achieving impedance matching to minimise reflected signals. Here we apply a microwave calibration technique to extract the S-parameters of a Josephson junction based TWPA in-operando. This enables reflections occurring at the TWPA and its extended network of components to be quantified, and we find that the in-operation performance can be well described by the off-state measured S-parameters. 
\end{abstract}
\maketitle

Quantum limited parametric amplifiers are becoming essential components in measurement chains for solid-state quantum devices and quantum computers. Recent years have seen tremendous advances in parametric amplifier technology \cite{macklin2015, esposito2021, malnou2024}, with a wide range of amplifier implementations \cite{eom2012, adamyan2016, phan2023, Faramarzi2024, qiu2023} and numerous commercial alternatives emerging. Of particular interest is broadband travelling-wave parametric amplifiers (TWPAs) as they offer great flexibility when operating at typical frequencies for quantum circuits; while still providing a sufficient amount of gain and SNR improvement for many applications. 

Due to their operation principles utilising propagating microwaves, TWPAs are very sensitive to their environment and the auxiliary components used in the setup. In particular, accurate impedance matching is crucial to avoid spurious reflected signals being amplified, resulting in gain ripples and reduced overall gain for the signal of interest \cite{kern2023, esposito2021, nilsson2024}. To this end, a refined knowledge of the detailed microwave performance (S-parameters) under different operating conditions will enable  further improvements in  amplifier performance. 
Previous amplifier developments critically focused on SNR improvement and noise performance, commonly utilising the Y-factor noise figure method \cite{Simbierowicz2021, malnou2023, ranadive2022, perelshtein2022}. A method for characterising device insertion loss relies on cold microwave switches and a separate thru-line which allows basic de-embedding of auxiliary circuitry such as coaxial wiring in the dilution refrigerator. More sophisticated implementations that rely on e.g. short-open-load (SOL) calibration standards \cite{Simbierowicz2022} have been demonstrated, however to achieve  small uncertainties in measurements this technique requires detailed knowledge of the mK performance of the standards used. In this context, an architecture based on a thru-reflect-line (TRL) calibration technique can more straightforwardly be used to obtain accurate calibration.

Here we demonstrate how to evaluate the two-port scattering parameters (S-parameters) of a commercial Josephson junction-based TWPA (JTWPA, Silent-Waves Argo \cite{planat2020}) during operation at mK temperatures using a low RF power TRL calibration technique compatible with quantum circuit operation \cite{stanley2022, stanley2022b, ranzani2013}.We also independently carry out a calibrated measurement of the auxiliary circuitry required to operate the JTWPA. In this way we obtain accurate S-parameter measurements for all the relevant driving conditions of the JTWPA. 
Our measurements can help inform improved impedance engineering as well as a detailed understanding of the impact from fabrication-induced parameter spread \cite{kissling2023, peatain2023} and external factors hampering TWPA performance \cite{nilsson2024}.

\begin{figure}
\centering
\includegraphics[scale=0.62]{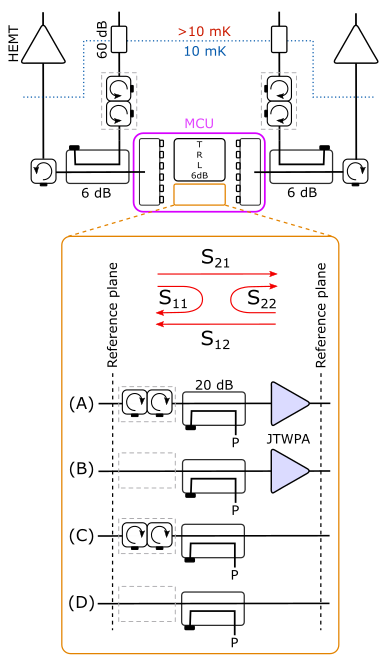}
\caption{\label{fig:schematic} Schematic of the mK S-parameter calibration  setup and the DUT networks measured. The calibration reference planes are moved to the ports of the 6-way cryogenic switches. Three ports are occupied by the TRL standards, at another port we mount a well-characterised 6 dB cryogenic attenuator for additional validation of the calibration, and on the two remaining switch ports we carried out measurements of the  JTWPA and it's auxiliary network as in configurations (A) \& (C) and (B) \& (D), in two separate cooldowns respectively. Dashed boxes indicate that isolators are either part of the DUTs or mounted outside the switch reference planes. }
\end{figure}

Non-idealities in the measurement setup caused by imperfect connectors and cabling will introduce errors in S-parameter measurements of the device under test (DUT). To measure the actual S-parameters at mK temperatures, a calibration scheme that shifts the reference planes to the input and output ports of the device is required, de-embedding the components between a room-temperature vector network analyzer (VNA) and the device at mK temperatures.

Our two-port S-parameter calibration setup has been specifically developed to characterise devices operating at very low power levels \cite{stanley2022} is shown in Fig. \ref{fig:schematic}, together with the four device configurations measured (A-D). The cold microwave calibration unit (MCU) consists of two 6-way cryogenic RF switches that are used to select between the TRL calibration standards or the DUTs, and they define the location of the calibrated reference planes. We have previously characterised the uncertainty introduced by these switches to be $<0.1$ dB in transmission at mK \cite{shin_broadband_2023}.
The setup utilises two heavily attenuated (50 dB) input lines and two output lines equipped with wideband (0.3-14 GHz) high electron mobility transistor (HEMT) amplifiers mounted on the 4 K stage of the cryostat. A 2-stage room temperature amplification chain further brings the signals to an acceptable level for VNA receiver measurements. 

The Thru standard is a zero-length insertable through connection of the nominally identical coaxial cables between switches and standards/DUT.
The Reflect standards are commercial (Maury Microwave 8046F6) 3.5 mm coaxial connectorized male and female offset short standards. 
The reference impedance of the calibration is the characteristic impedance of the Line standard, which has been measured to be very close to 50 $\Omega$ ($49.94\pm0.03$ $\Omega$ in the frequency range 2-8 GHz) and is temperature invariant from 25 mK to 296 K \cite{skinner2023}. We have previously characterised the error budget of our setup \cite{skinner2023, shin_broadband_2023, stanley_four_port_2024}, finding a reflection coefficient uncertainty of about 0.04 in linear units.

All four uncalibrated S-parameters are obtained by measuring the respective RF input and output coaxial lines which connects the MCU to a 4-port VNA (PNA-X N5247B).
The JTWPA pump line (indicated 'P' in Fig. \ref{fig:schematic}) is configured with 6/10/10/6/10 dB attenuation at 50K/4K/800mK/100mK/10mK stages of the fridge respectively. For all the lines we also use 0.25-10 GHz band-pass filters (not drawn) at the 10 mK stage.

Impedance matching and suppression of reflections are essential for good amplifier performance, and it is common practice to place isolators both before and after the TWPA. In the former case it also protects a qubit sample from backaction due to pump leakage. In all cases we use a single junction isolator on the output lines on the common port of the RF switches. On the input side of the JTWPA we use a double junction isolator as indicated in Fig. \ref{fig:schematic} with grey dashed boxes, either as part of the DUT (cases A and C) or as part of the common input line (cases B and D), the data of which was collected in two consecutive cooldowns. In all cases the cables used in-between the components were ensured to be of the same length and type. In this work we perform all the cold stage measurements in the frequency range 4-8 GHz, limited by the cryogenic isolators used. One complication for S-parameter measurements in this architecture is that if the isolators are placed inside the calibration reference planes, only limited information about the JTWPA can be obtained. On the other hand, if placed further from the JTWPA there is a chance of introducing additional reflections that can be amplified as a result of the almost unitary reverse transmission of the JTWPA. Therefore, an understanding of reflections occurring in the wider network of components is also crucial.

\begin{figure}
\centering
\includegraphics[scale=0.52]{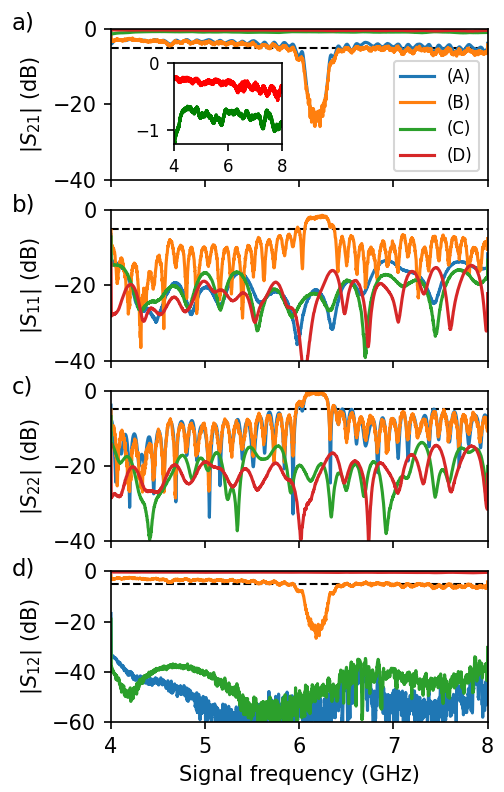}
\caption{\label{fig:abcd} S-parameters for configurations  (A), (B), (C), and (D) obtained at 10 mK and typical DUT input signal levels of $\approx-110$ dBm. The inset in a) shows a close-up of the measured $S_{21}$ response of the coupler and cables, and isolator, configurations (D) and (C) respectively. In all cases the JTWPA pump tone is off.}
\end{figure}

To extract the actual S-parameters of the DUT at mK temperatures, we solve the 8-term error model \cite{engen1979} applied to the uncalibrated S-parameters measured with the VNA. This model accounts for systematic errors such as directivity, source match and reflection tracking which are generated due to reflections in the measurement setup. 
To validate the calibration process \cite{stanley2023} we also used the aforementioned cryogenic 6 dB attenuator as DUT in each cooldown. 
For all the measurements (unless otherwise mentioned) we use a low input power (-30 dBm on the VNA, $< -110$ dBm incident on the JTWPA) to ensure we are operating well below JTWPA gain compression.

In Fig. \ref{fig:abcd}, we show the measured S-parameters of all four configurations shown in the schematic of Fig. \ref{fig:schematic}.
We clearly see that the insertion and return losses due to the directional coupler (used to inject the pump tone) and the cables connecting it to the JTWPA are small ($<1$ dB and $\lesssim -20$ dB respectively; configuration D), and so we can neglect these components without affecting the conclusions of the JTWPA measurement. Furthermore, we clearly suppress $S_{11}$ (Fig. \ref{fig:abcd}b) and reverse transmission (Fig. \ref{fig:abcd}d) by inclusion of the isolator before the coupler (configuration (A) and (C)). Thus we can conclude that when we measure configuration (B) the response accurately represents the JTWPA performance alone.

When the pump tone is off, we measure an insertion loss of the JTWPA varying from 3 to 6 dB across the 4-8 GHz frequency range (Fig. \ref{fig:abcd}a). 
The reflection measured at the two ports ($S_{11}$, Fig. \ref{fig:abcd}b, and $S_{22}$, Fig. \ref{fig:abcd}c) remains near or below -10 dB, indicating a device closely matched to $50$ $\Omega$. 

\begin{figure}
\centering
\includegraphics[width=0.50\textwidth]{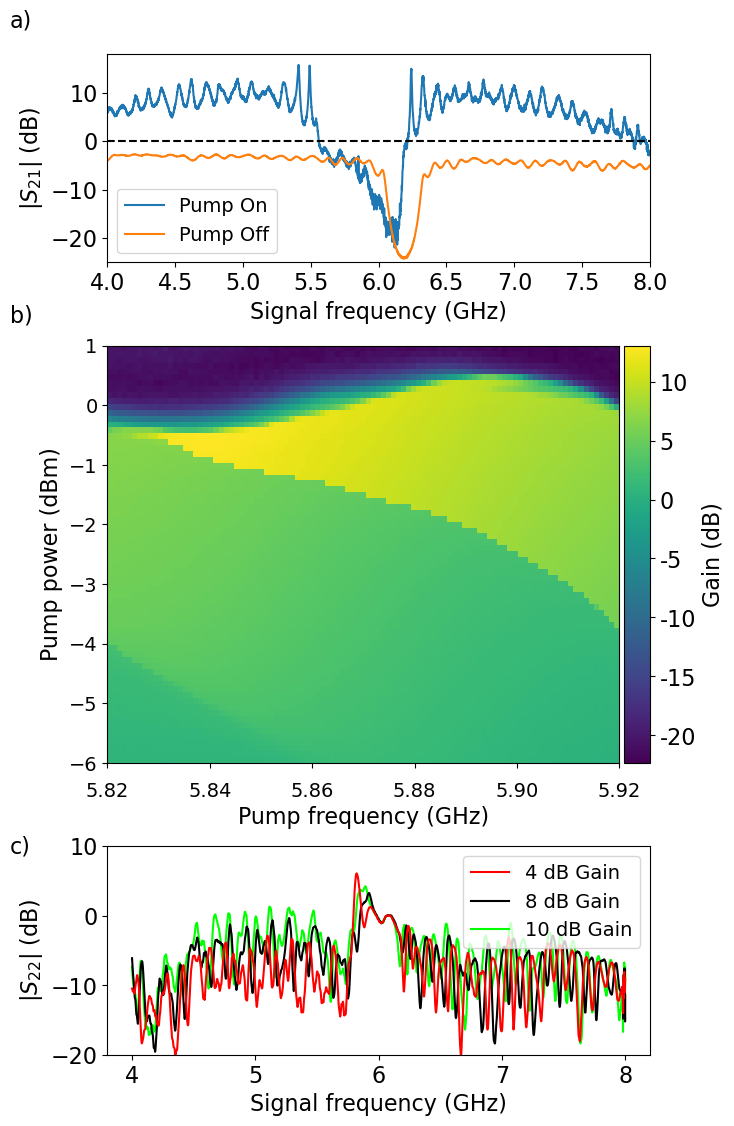}
\caption{\label{fig:b} {{In-operation performance.}} a) Example $S_{21}$ magnitude with the pump on and off. Dashed lines shows 0 dB as a reference. b) Average gain as a function of pump power and frequency.  c) Typical examples of $S_{22}$ at selected gains of $4$ dB (taken at $P_p=-4.3$ dB, $f_p=5.835$ GHz), $8$ dB ($-1.5$ dB, $5.02$ GHz), and $10$ dB ($-1.5$ dB, $5.875$ GHz).  A moving average is used to better illustrate the overall trend of increasing $S_{22}$ magnitude with increasing gain.}
\end{figure}

\begin{figure}
\centering
\includegraphics[width=0.38\textwidth]{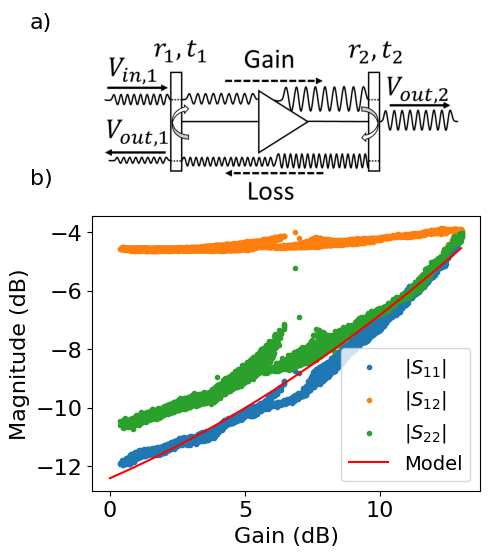}
\caption{\label{fig:c} {{S-parameters in the presence of gain.}} a) Sketch of the setup used to model the role of reflected signals on the measured S-parameters of the JTWPA, see text for details. b)  Average S-parameter magnitude as a function of gain. The solid line indicates the expected trend assuming the measured average S-parameters of the JTWPA pump off state and the model in a). }
\end{figure}

Next, we turn on the JTWPA pump tone.
In Fig. \ref{fig:b}a we show an example of the measured $S_{21}$ magnitude with the pump signal on and pump off. In what follows we define the gain as the difference between $S_{21}$ with the pump on and pump off. This should be compared to the 'useful' gain of the JTWPA relative to the case of no JTWPA at all, i.e. within our calibrated setup referenced to 0 dB.
In Fig. \ref{fig:b}b we show the gain as a function of the pump frequency and power.  The gain is averaged across all signal frequencies in the range 4-8 GHz, excluding the stop-band ($5.5$ GHz to $6.5$ GHz).   
In Fig. \ref{fig:b}c we show selected $S_{22}$ traces for different gain as a function of signal frequency, showing a clear overall increase in $S_{22}$ with increasing gain. To quantify this in more detail we show in Fig. \ref{fig:c}b the change in the similarly averaged scattering parameters as a function of gain.
We see a significant increase in $S_{11}$ and $S_{22}$, suggesting increased reflection from the device.  This naively suggests that the impedance of the JTWPA changes with gain, however, as we will see this is a result of the initial impedance mismatch of the circuit, and the amplification of reflected signals. In the extreme cases this gain results in reflected S-parameters exceeding $0$ dB (Fig. \ref{fig:b}b). The impedance of the JTWPA itself is expected to change only by a very small amount under these pump conditions \cite{planat2020}.  

A simple model (sketched in Fig. \ref{fig:c}a) can be used to investigate the behaviour of $S_{11}$ and $S_{22}$ with increasing gain: similarly to the Fabry-Perot cavity model described in \cite{planat2020}, the system can be considered as two input/output ports with linear reflection coefficients $r_1$ and $r_2$ and transmissions $t_1$ and $t_2$, with $[t_{i}^2 + r_{i}^2 = 1 ]_{i = 1,2}$ in the lossless case. An incoming signal arrives at port 1 with a proportion $r_1$ being reflected back to the source and $t_1$ entering the amplifier. From here the signal is amplified by the gain coefficient $g = \sqrt{G}$, with $G$ the measured power gain, before being partially transmitted out of the amplifier through port 2, and partially reflected to stay within the amplifier. The signal continues to reflect back and forth within the amplifier, with a proportion transmitted at each port each time, whilst also being amplified between ports 1 and 2. The resulting ratio of the amplitudes of the output and input signal voltages from the arising geometric series can be expressed as:
\begin{equation}
\frac{V_{out}}{V_{in}} = r_1 + \frac{t_1^2gr_2}{1-gr_2r_1},
\label{eqn:s11_model}
\end{equation}
with the corresponding model for a signal arriving at port 2 obtained by swapping the subscripts. This model was used to calculate the average reflection coefficients from the pump-off $S_{11}$ and $S_{22}$ data in Fig. \ref{fig:abcd}, assuming return loss of 3.5 dB as seen in the $S_{21}$ data, producing values of $r_1 = r_2 \approx 0.14$. These reflection coefficients are used in Eq. \ref{eqn:s11_model} to calculate the expected change in $S_{11}$ and $S_{22}$ with increasing gain, which is plotted in Fig. \ref{fig:c}b and shows agreement within the measurement error.

This model and measurement is not able to distinguish reflections occurring directly at the JTWPA ports from reflections occurring further from the amplifier or even outside the reference planes for calibration. However, we can neglect reflections due to other components within the measurement reference plane as $S_{11}$ and $S_{22}$ for cases C and D are well below -20 dB. Furthermore, inclusion of the isolator within the reference planes (case A) would eliminate any reflections occurring outside the reference plane before the JTWPA and thus set $r_1\approx 0$. Yet, in this case we still observe the same behaviour of the measured reflection vs gain. 

\begin{figure}
\centering
\includegraphics[width=0.43\textwidth]{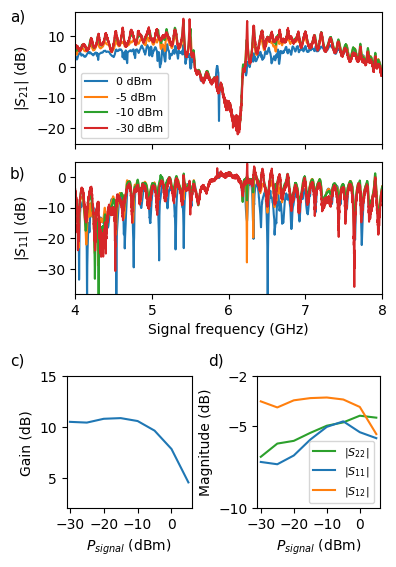}
\caption{\label{fig:psat} {{Performance near gain compression.}} a) $S_{21}$ and b) $S_{11}$ magnitude vs signal frequency as a function of signal power (referenced to VNA output) taken with the pump tuned close to maximum gain ($f_p=5.8659$ GHz, $P_p =-0.7$ dBm).  
 c) The average gain in the band 4.5-5.5 and 6.5-7.5 GHz as a function of signal power (referenced to the generator output level). d) Average change in S-parameters as a function of signal power evaluated in the same frequency band. }
\end{figure}

All previous data was taken with a very low input signal power to the DUT ($\approx -110$ dBm) to ensure measurements were performed without saturating the JTWPA. As the last step we characterise the response of the JTWPA as we increase the input signal power and start to observe gain compression. The magnitude of the S-parameters for a number of different signal powers are shown in Fig. \ref{fig:psat}, taken at a point near maximum gain ($\sim 11$ dB; $f_p=5.8659$ GHz, $P_p =-0.7$ dBm). These measurements confirm that previous measurements were done with a sufficiently low signal power to avoid any effects due to saturation. As the overall gain is suppressed by the signal power we observe a non-trivial dependence of the reflection at the two ports (Fig. \ref{fig:psat}d), uncorrelated with the gain suppression (Fig. \ref{fig:psat}c). Furthermore, $S_{12}$ also drops sharply when the gain is significantly suppressed. 
Together, this indicates that the signal saturation results in changes to the devices intrinsic dissipation. 

Measurements of S-parameters of a TWPA presents a challenge due to its non-linear and near-reciprocal response. Ideal operation seeks to minimise reflections utilising isolators close to each port of the TWPA, however, inclusion of isolators obscures the TWPA response in calibrated measurements. Future calibrations methods would benefit from more advanced techniques such as also measuring the absolute power incident on the two ports \cite{honigl-decrinis_two-level_2020}, or X-parameters and large-signal analysis \cite{roblin}.

In summary, we have performed in-situ, in-operando microwave S-parameter measurements of a JTWPA and its auxiliary network of components in a calibrated setup. We reveal how the S-parameters of the JTWPA depend on the strength of the pump and signal power, allowing us to understand how reflections influence JTWPA performance, and show how the JTWPA off-state S-parameters can accurately describe the on-state behaviour. Our method allows to develop detailed models of the device physics based on the observed device characteristics, and fine-tune parametric amplifier design to improve performance. 

We acknowledge fruitful discussions with Luca Planat and Silent Waves. We acknowledge the support from the UK Department for Science, Innovation and Technology through the UK National Quantum Technologies Programme (NQTP). We also acknowledge support from the Engineering and Physical Sciences Research Council (EPSRC) (Grant Number EP/W027526/1).

\bibliography{sample}

\end{document}